\documentclass[aps,pre,preprint,superscriptaddress]{revtex4}

\usepackage{graphicx}
\usepackage{graphics}
\usepackage{epsfig}
\usepackage{amssymb}
\usepackage{amsmath}
\usepackage{color}
\usepackage{textcomp}
\usepackage[utf8]{inputenc}
\usepackage{bm}
\usepackage{tabularx}

\begin{document}

\title{Coarsening Dynamics in the Vicsek Model of Active Matter}

\author{Nisha Katyal}
\affiliation{School of Physical Sciences, Jawaharlal Nehru University, New Delhi- 110067, India.}
\author{Supravat Dey}
\affiliation{Laboratoire Charles Coulomb Universit\'e Montpellier and CNRS, UMR 5221, 34095 Montpellier, France.}
\author{Dibyendu Das}
\affiliation{Department of Physics, Indian Institute of Technology Bombay, Powai, Mumbai 400076, India.}
\author{Sanjay Puri}
\affiliation{School of Physical Sciences, Jawaharlal Nehru University, New Delhi- 110067, India.}

%\begin{center}
%{\Large{\bf Coarsening Dynamics in the Vicsek Model of Active Matter}} \\
%\ \\
%\ \\
%by \\
%Nisha Katyal$^{1}$, Supravat Dey$^{2}$, Dibyendu Das$^{3}$ and Sanjay Puri$^{1}$ \\
%$^1$School of Physical Sciences, Jawaharlal Nehru University, New Delhi -- 110067, India. \\
%$^2$Laboratoire Charles Coulomb Universit\'e Montpellier and CNRS, UMR 5221, 34095 Montpellier, France. \\
%$^3$Department of Physics, Indian Institute of Technology Bombay, Powai, Mumbai 400076, India.
%\end{center}

\begin{abstract}
We study the flocking model introduced by Vicsek \emph{et al.} (1995) \cite{vic95} in the ``coarsening" regime. At standard self-propulsion speeds, we find two distinct growth laws for the coupled density and velocity fields. The characteristic length scale of the density domains grows as $L_{\rho}(t) \sim t^{\theta_\rho}$ (with $\theta_\rho \simeq 0.25$),  while the velocity length scale grows much faster, $viz.$, $L_{v}(t) \sim  t^{\theta_v}$ (with $\theta_v \simeq 0.83$). The spatial fluctuations in the density and velocity fields are studied by calculating the two-point correlation function and the structure factor, which show deviations from the well-known Porod's law. This is a natural consequence of scattering from irregular morphologies that dynamically arise in the system. At large values of the scaled wave-vector, the scaled structure factors for the density and velocity fields decay with powers $-2.6$ and $-1.52$, respectively.
\end{abstract}
\maketitle

%\newpage
\section{Introduction}

The nonequilibrium coarsening dynamics of statistical systems towards their steady states has been of  long-standing interest \cite{bray,pw09}.  A classic example is a quenched Ising ferromagnet at  temperature $T<T_c$,  which approaches the ordered equilibrium state by coarsening of up/down spin domains.  The evolving morphology is statistically self-similar in time, provided spatial distances are scaled by a unique  time-dependent length scale called the `coarsening length'  $L(t)$. Typically $L(t)$ has a power-law dependence, i.e., $L(t) \sim t^{\theta}$.  For pure and isotropic ferromagnetic systems, the growth exponent $\theta=1/2$ for nonconserved spin-flip kinetics, and $1/3$ for conserved spin-exchange kinetics \cite{pw09}.  Even for systems with nonequilibrium steady states (which violate detailed balance in contrast to equilibrium systems like Ising model), coarsening behaviour from a homogeneous to an increasingly clustered state has been of great interest.  For example, logarithmic dependence of $L(t)$ was found for certain driven lattice models \cite{evans,rama2000}, while the more common power law form of $L(t)$ was found for various others --- particles sliding on fluctuating surfaces \cite{das,manoj03}, freely cooling granular gases \cite{dp03,ap06,shinde}, and models of active matter \cite{supravat,hagan2013,beatrici17,dey19}.  In this paper, we study in detail the coarsening kinetics in a non-equilibrium system of self-propelled particles, namely the Vicsek model \cite{vic95}. Most previous studies focused on the steady-state properties of this model and its coarse-grained counterparts. In contrast, we focus on the ordering kinetics from a homogeneous initial condition evolving towards the steady state. Contrary to most coarsening systems mentioned above, which usually have a single dominant length scale characterising  the growth of order, we show in this work that Vicsek model is unusual --- there are two distinct coarsening length scales governing the dynamical behaviour of the particle density and velocity fields. 

Studies of ``active matter" are of great contemporary interest in statistical physics \cite{vicsek2012,RamaRev}. The Vicsek model \cite{vic95} belonging to this field, is a pioneering one representing systems of self-propelled particles with polar degrees of freedom.  Self-propulsion and alignment with neighbours, give rise to macroscopic velocity ordering in $2$-dimensions. Several variants  of the  model with modified update rules, vectorial instead of angular noise, and additional short-range interactions have been studied \cite{gregorie04,chate08}--- they clarified the nature of the noise-driven non-equilibrium phase transition and also showed new features like travelling bands. Extensive studies of the continuum limit of the system was done analytically \cite{toner95,tu98,toner2005,bretin06,mishraPRE10}, and partly numerically \cite{mishraPRE10}. An experimental realisation of the Vicsek model was attained in a laboratory system of vibrated asymmetric granular discs \cite{deseigne10}. Although a natural expectation maybe that the model represents a bird flock, systematic studies of starling bird flocks \cite{cavagna08,cavagna14} showed that unlike the Vicsek model, the interactions of birds are `topological' --- a modified model was developed in this context \cite{chate_topo10}. Recently, in binary mixtures, a Vicsek-like alignment interaction was incorporated and was seen to facilitate phase separation \cite{subir2014}. We would like to note that active matter systems are varied and complex; objects with nematic symmetry form another big domain of interest \cite{zhang,marchetti_rev2013,mishra14}. 
 
There have been very few studies of coarsening in active systems. A numerical study of the continuum equations for polar models was done by Mishra \emph{et al.} \cite{mishraPRE10}. The density structure factors $S_{\rho \rho}(\vec{k}, t)$ and associated giant number fluctuations were studied by Dey \emph{et al.} \cite{supravat} in the coarsening regime of various active models. The $S_{\rho \rho}(\vec{k}, t)$ for Vicsek-like models showed an interesting crossover behaviour--- the small wave-vector scaling was generic and related to giant number fluctuations, while the large wave-vector scaling indicated Porod law \cite{porod} violations. Furthermore, a coarsening length scale for density, namely $L_{\rho}(t) \sim t^{\theta_\rho}$ (with $\theta_\rho \simeq 0.25$), was reported for the Vicsek and other polar models with local alignment rules \cite{supravat}.  Interesting recent studies of active discs without Vicsek-like alignment rules (and hence showing no macroscopic velocity order) \cite{hagan2013,cates2013,beatrici17, dey19}, reported similar coarsening exponents $\theta_\rho \simeq 0.23-0.28$ for the density. In an entirely different context, coarsening dynamics of passive advective density field in an active field of dynamic asters was recently studied \cite{madan}.
 
To the best of our knowledge, study of coarsening of the velocity field in Vicsek-like models has been rare. In this paper, we undertake a comprehensive study of the coarsening problem in the Vicsek model for both the velocity and density fields. A natural question is that, as the particles cluster together and form density domains, do their velocities align over the same length scale? As we will show below, the velocities align over a distinct  and faster growing length scale $L_v(t) \sim t^{\theta_v}$ (with $\theta_v \simeq 0.83$), as  compared to $L_{\rho}(t) \sim t^{\theta_\rho}$ (with $\theta_\rho \simeq 0.25$).  The value of $\theta_v$ seems like the inverse of the dynamical exponent $z = 6/5$ predicted from continuum theory in the steady state \cite{toner2005}. We study various statistical quantities to establish these facts below.

This paper is organized as follows. In Sec.~\ref{s2}, we introduce the Vicsek model and present its coarsening snapshots. In Sec.~\ref{s3}, we discuss the quantities used to characterize the coarsening process. In Sec.~\ref{s4}, we present detailed numerical results for these quantities. The paper concludes with a summary and discussion in Sec.~\ref{s5}.

\begin{figure}[h]
\centering
\includegraphics[width=\textwidth]{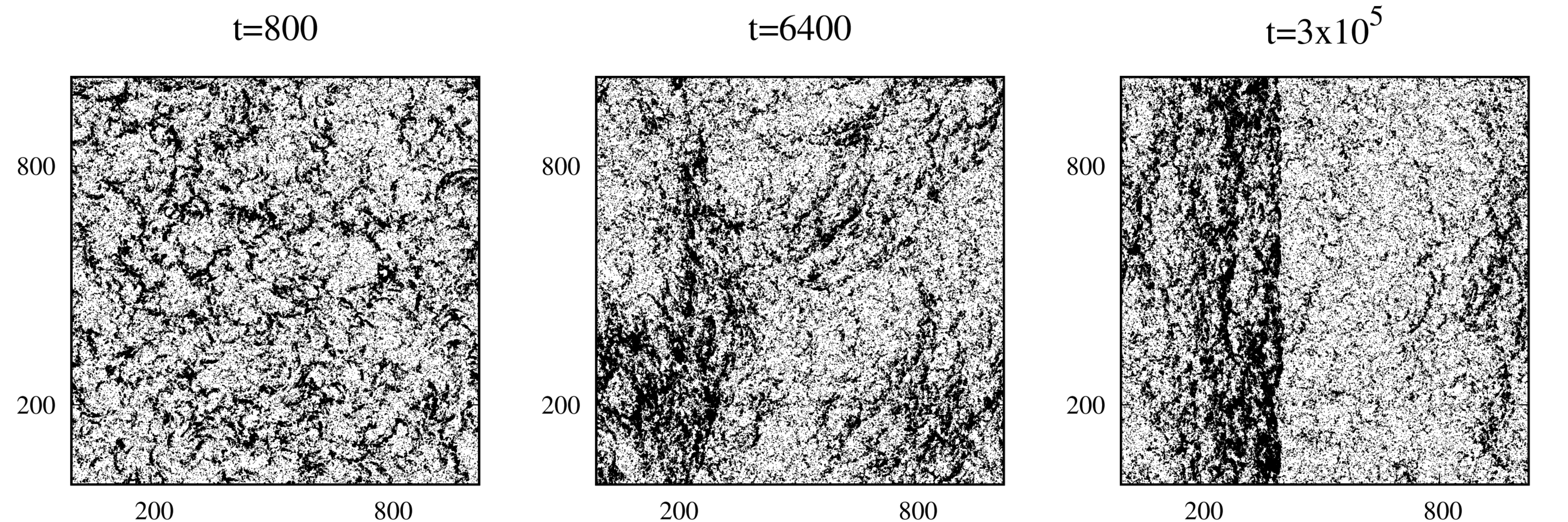}
\caption{Snapshots for $t=800, 6400$, and $3\times10^5$. The box length is $L=1024$; all other parameters are defined in the text. The location of the particles is marked in black, while the white regions denote empty spaces. The left two figures are in the early coarsening regime, while the right one where a clear density band is visible is in a late stage approaching the steady state.}
\label{fig1}
\end{figure}

\section{Vicsek Model and Coarsening Snapshots}
\label{s2}

The Vicsek model (VM) in $d=2$, has the following equations of motion for the position ${\bf r}_i$ and velocity ${\bf v}_i =  (v_0 \cos \theta_i, v_0 \sin \theta_i)$ of the $i$-th particle \cite{vic95}:
\begin{eqnarray}
	{\bf r}_{i}(t+\Delta t)={\bf r}_{i}(t)+{\bf v}_{i}(t) \Delta t, \label{pos} \label{position}\\
\theta_{i}(t+\Delta t)=\arg\left [\sum_{k} \exp(i\theta_{k}(t))\right ]+\eta \xi(i,t) \label{theta}.
\label{eqn:update}
\end{eqnarray}
An average alignment direction is obtained by summation over neighbouring ``$k$"-particles around particle $i$ (including itself), over a circle of radius $R$. The second term to the right of Eq.~(\ref{theta}) denotes the errors made in alignment along the latter average direction--- a random angle is added lying uniformly between $-\eta \pi$ to $\eta \pi$. Note that $\eta$ is a fixed chosen number $\in [0,1]$, while $\xi(i,t)$ is a random number drawn from a uniform distribution between $[-\pi, \pi]$.

We choose a simulation box of size $L \times L$ (with periodic boundary conditions) and $L$ ranging from $512$ to $4096$. The number of particles $N=\rho_{0}L^{2}$, and we choose the average particle density $\rho_{0}=1$. The update time interval $\Delta t = 1$, the noise amplitude is fixed at $\eta=0.3$, and the speed $v_0 = 0.5$. Initially the system has spatially (uniformly) randomly distributed particles, which have (uniformly) randomly oriented velocity vectors. For the above mentioned parameters, the Vicsek model has a transition to a polar ordered state below a critical noise strength $\eta_c \simeq 0.45$ (numerically known).  At $\eta = 0.3$, the velocity order parameter attains a value $\approx 0.65$ in the steady state for the system sizes we study (see below), and hence the system coarsens towards an ordered state. The statistical properties of variants of the Vicsek model have been a matter of intense debate \cite{vicsek2012,nagy07,albano08,albano09,Biglietto2012,gregorie04,chate08,chate15, shaebani19}. Recent works show that density bands are expected generically near the steady state \cite{vicsek2012,chate15}. Our study focuses on the time-dependent behavior in coarsening, far from the steady state, for the original Vicsek model with the update rules defined in Eqs.~(\ref{position}) and (\ref{theta}).

Fig.~\ref{fig1} shows the density coarsening for the Vicsek model, for $L=1024$ and the parameters mentioned above. Up to times $t \sim 10^4$, for which we present all our coarsening data below, we found that there is no anisotropy due to spatial band formation.  However, the density bands do appear at late times ($\sim 10^5$) approaching the steady state (see Fig.~\ref{fig1}).  For our numerical  studies up to $t \sim 10^4$, the absence of density bands ensure that we may do isotropic averaging of the correlation functions. We now proceed to define the relevant correlation functions.

\section{Characterization of Coarsening Morphologies}
\label{s3}

We define a local coarse-grained box density $\rho({\vec r},t)$ as the total number of particles in a square box of side $b$. Similarly a local coarse-grained box velocity ${\bf v}({\vec r},t)$ is the average velocity over all particles contained in a box. We have used $b=1$. To study the morphology of spatial density structures, we use the equal-time density correlation function and its Fourier transform (the structure factor) \cite{pw09}:
\begin{eqnarray}
C_{\rho \rho}({\vec r},t)&=&\langle \rho({0},t)\rho({\vec r},t) \rangle - \langle \rho({0},t)\rangle \langle\rho({\vec r},t) \rangle, \\
S_{\rho \rho}({\vec k},t)&=& \int_{-\infty}^{\infty}C_{\rho \rho}({\vec r},t) e^{i {\vec k} \cdot {\vec r}} d{\vec r} = \langle \tilde{\rho}({\vec k},t)\tilde{\rho}({-\vec k},t) \rangle.
\end{eqnarray}
Here, $\langle...\rangle$ denotes averaging over independent initial conditions (typically $\sim 100$ in our simulations), and 
$\tilde{\rho}$ is the Fourier transform of $\rho$. Similarly to study the correlations in polar alignment, we use the velocity-velocity correlation function and the corresponding structure factor:
\begin{eqnarray}
C_{vv}({\vec r},t) &=& \langle {\bf v}({0},t)\cdot {\bf v}({\vec r},t) \rangle - \langle {\bf v}({0},t) \rangle \cdot \langle{\bf v} ({\vec r},t) \rangle, \\
 S_{vv}({\vec  k},t) &=& \int_{-\infty}^{\infty} C_{vv}({\vec r},t) e^{i {\vec k} \cdot {\vec r}} d{\vec r}.
\end{eqnarray}
As shown in snapshots of the VM model in the previous section there is no anisotropic density structure up to times $t \sim 10^4$. Hence using the circular symmetry, the correlation functions $C_{\rho \rho}({\vec r},t)$ and $C_{vv}({\vec r},t)$ and structure factors $S_{\rho \rho}({\vec k},t)$ and $S_{vv}({\vec  k},t)$ are circularly averaged over \emph{all} orientations of $\vec r$ and $\vec k$, respectively. Our results below are shown as a function of the radial distance $r = |{\vec r}|$ and wave-vector magnitude $k = |{\vec k}|$.

Before presenting our results, we make some general observations. The correlation function and structure factor typically have the following scaling forms \cite{pw09}, when distance or wave-vector is 
scaled by the coarsening length $L(t)$:
\begin{eqnarray}
C(r,t) &=& g(r/L(t)), \label{Cr}\\
S(k,t) &=& L(t)^d f(k L(t)).
\end{eqnarray}
For scalar order parameters like the density field, the short-distance behavior of the scaling function $g(x)$ is 
\begin{eqnarray}
g(x) =  1 - A\,x^{\alpha} + \cdots,
\label{cusp}
\end{eqnarray}    
valid for $a \ll r \ll L(t)$, where $a$ is the microscopic scale. If the domains have smooth boundaries, and inter-domain separations have no hierarchy of length scales, $\alpha=1$ indicating the well-known Porod decay \cite{porod,bray, pw09,puri88}. Interestingly, many cases of Porod law violation are also known in the literature, where $\alpha<1$. This may arise when domains have fractal surface or volume morphologies \cite{bale,gaurav,shrivastav11, shrivastav14,bupathy16}, or inter-cluster separations are hierarchic obeying power-law distributions \cite{das,shinde,s_mishra06,supravat,anu_rajesh}. The corresponding large-$k$ behavior of the scaling function of structure factor is
\begin{eqnarray}
f(k) &=& \tilde{A} \, k^{-\left(d+\alpha\right)} + \cdots, 
\end{eqnarray} 
valid for $1/a \gg k \gg 1/L(t)$.  The Porod law corresponds to $f(k) \sim k^{-\left(d+1\right)}$. 

In contrast to this, for a vector order parameter with $n$ components, the structure factor exhibits the generalized Porod or Bray-Puri-Toyoki \cite{puri91,toyoki,bray} (BPT) tail:
\begin{eqnarray}
f(k) &=& \tilde{A}_n \, k^{-\left(d+n\right)}.
\end{eqnarray} 
This is a consequence of scattering from vector defects like vortices ($n=2$), monopoles ($n=3$), etc.. The scalar result is recovered for $n=1$, corresponding to the case of interface defects. For the velocity field in the Vicsek model (in $d=2$), we have $n=2$ and would expect a BPT tail $f(k)\sim k^{-4}$. To the best of our knowledge, there are no reported violations of the BPT tail for vector fields. We will see an exception below.

\section{Numerical Results}
\label{s4}

\begin{figure}[!ht]
\vspace{0.7cm}
\centering
\includegraphics[width=\textwidth]{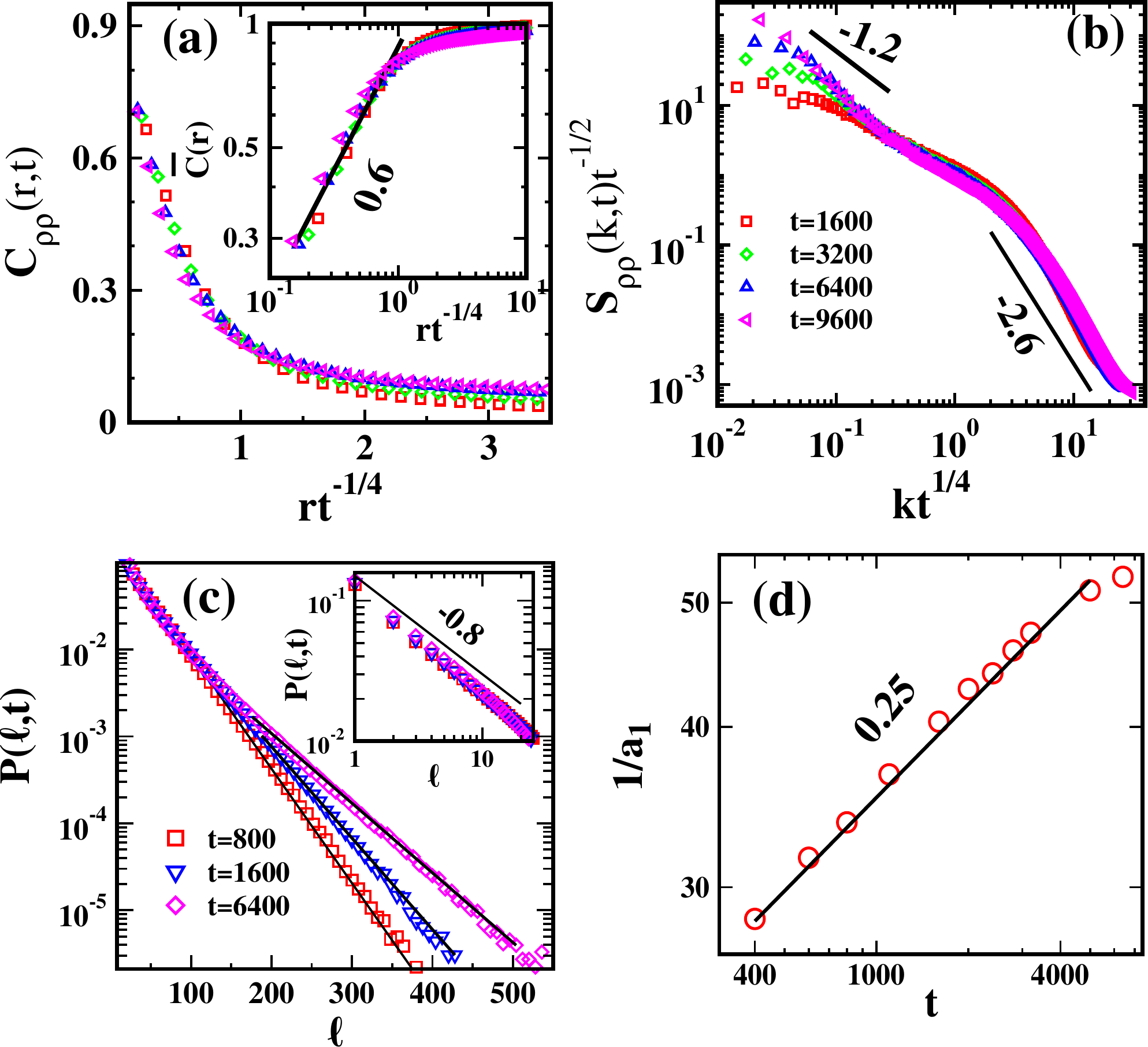}
\caption{(a) Scaled density-density correlation function $C_{\rho \rho}(r,t)$ (for times $t=1600, 3200, 6400$ and $9600$) obtained using a coarsening length $L_{\rho}(t) \sim t^{0.25}$. Inset shows a log-log plot of ${\bar{C}}$ (see text) -- the power law exponent $\alpha=0.6$ is shown. (b) Scaled structure factor decay crosses over from a power-law with exponents -1.2 [for small $kL_{\rho}(t)$] to -2.6 [for large $kL_{\rho}(t)$]. (c) Probability distribution function $P(\ell,t)$ as a function of domain width $\ell$ in a linear-log plot showing a power law behaviour (for small $\ell$) cut-off by an exponential (at large $\ell$). For clarity, the power-law behaviour for small $\ell$, is shown in the inset.  (d) A log-log plot of variation of the inverse of decay length $1/a_{1}$ (characteristic cluster width) as function of time $t$. A power law $\sim t^{0.25}$ is put against the data.}
\label{fig2}
\end{figure}

\subsection{Ordering of Density Field}
\label{s41}

At first we study the density ordering, elaborate beyond \cite{supravat} by presenting new results, and clarify various subtle points. In Fig.~\ref{fig2}(a), we show  $C_{\rho \rho} (r,t)$ vs. $r/{L_{\rho}(t)}$ for  different times. We assume a power law form for $L_{\rho}\sim t ^{\theta_\rho}$, and adjust $\theta_\rho$ to get a collapse of the data. We find that $\theta_\rho \simeq 0.25$ collapses the data for small and intermediate $r/{L_{\rho}(t)}$ quite well, but  fails at large scaled distances. This is consistent with the data collapse of scaled $S_{\rho\rho}(k,t)$ over two decades of $kL_{\rho}(t)$ (at large and intermediate wave vectors);  the collapse is poor for small $kL_{\rho}(t)$. We would investigate more critically the validity of the exponent $\theta_\rho \simeq 0.25$ below, but before that, let us discuss the structural information conveyed by the two point functions.

In the inset of Fig.~\ref{fig2}(a), we plot $\bar C(r)=1-C(r,t)$ vs. $r/L_{\rho}$. This plot shows a power-law behaviour with a cusp exponent $\alpha=0.6$ (see Eq.~\ref{cusp}) for $r/L_{\rho}\ll1$, indicating a Porod law violation. In accordance with this, in Fig.~\ref{fig2}(b), we see a power law decay in $S_{\rho\rho}$ with a power $-(d+\alpha) = -2.6$ (for large $kL_{\rho}$).  This violation of the Porod law underlines a remarkable fact -- the density clusters for VM have an irregular fractal morphology associated with them.  Another distinctive feature of the data in Fig.~\ref{fig2}(b) (which is a characteristic of other polar Vicsek-like models too \cite{supravat}) is that it has two different power laws at small and large wave vectors.  The small $k$ power law is hard to conclude from the data we have, but we expect it to be approaching a power $-1.2$, for the following reasons. The number fluctuations $\sigma_l^2 = \langle (N - \langle N \rangle)^2 \rangle_l$  in a $l \times l$ box, and the density structure factor $S_{\rho\rho}(k)$ are related as $\sigma_l^2 = l^d S_{\rho \rho} (k\to 0)$. For active systems, the number fluctuations are often ``giant'' violating the central limit theorem, i.e., $\sigma_l^2 \sim \langle N \rangle_l^{\beta}$ with $\beta>1$. If $S_{\rho\rho}(k) \sim k^{-(d-\eta)}$ as $k \rightarrow 0$, putting $k \sim 1/l$ and noting $\langle N \rangle \sim l^d$, it follows that $\sigma_l^2 \sim \langle N \rangle_l^{2 - \eta/d}$.  For VM in the coarsening regime, our earlier work had shown that $\beta = 1.6$ \cite{supravat}. This value of $\beta$ implies a value of the power of $d - \eta = 1.2$ for  the density structure factor at small $k$. Apart from that in continuum theories this power $-1.2$ is known to appear in the structure factor at small $k$ \cite{tu98}.

To understand the origin of $\theta_{\rho} \simeq 0.25$, we study the probability distribution $P(\ell,t)$ of coarse-grained domain widths $\ell$. A way of marking coarse-grained domains is to put ``spin'' variables $n({\vec r}) = +1$ or $-1$, respectively, if local box density $\rho({\vec r}) >\rho_0$ or $\rho({\vec r}) <\rho_0$. Scanning the  system horizontally and vertically for continuous spatial stretches of $n({\vec r}) = -1$ provide samples of length $\ell$. The distributions $P(\ell,t)$ vs. $\ell$ at different times, are shown in Fig.~\ref{fig2}(c). For small $\ell$, there is a power-law behavior with exponent $-0.8$, as shown in the inset. For large $\ell$, clear exponential tails can be seen in the log-linear plot of the main Fig.~\ref{fig2}(c). Fitting the tails to $\sim \exp(-a_1 \ell)$, we extract $a_1$ for various times $t$. The inverse of the decay constants are plotted in Fig.~\ref{fig2}(d), and we see a clear power law $1/a_1 \sim t^{0.25}$.  This lends credence to the conclusion that  lengthscale $L_{\rho}(t) \sim t^{0.25}$ is directly related to the size of the growing density clusters.

\begin{figure}[!ht]
\centering
\includegraphics[width=\textwidth]{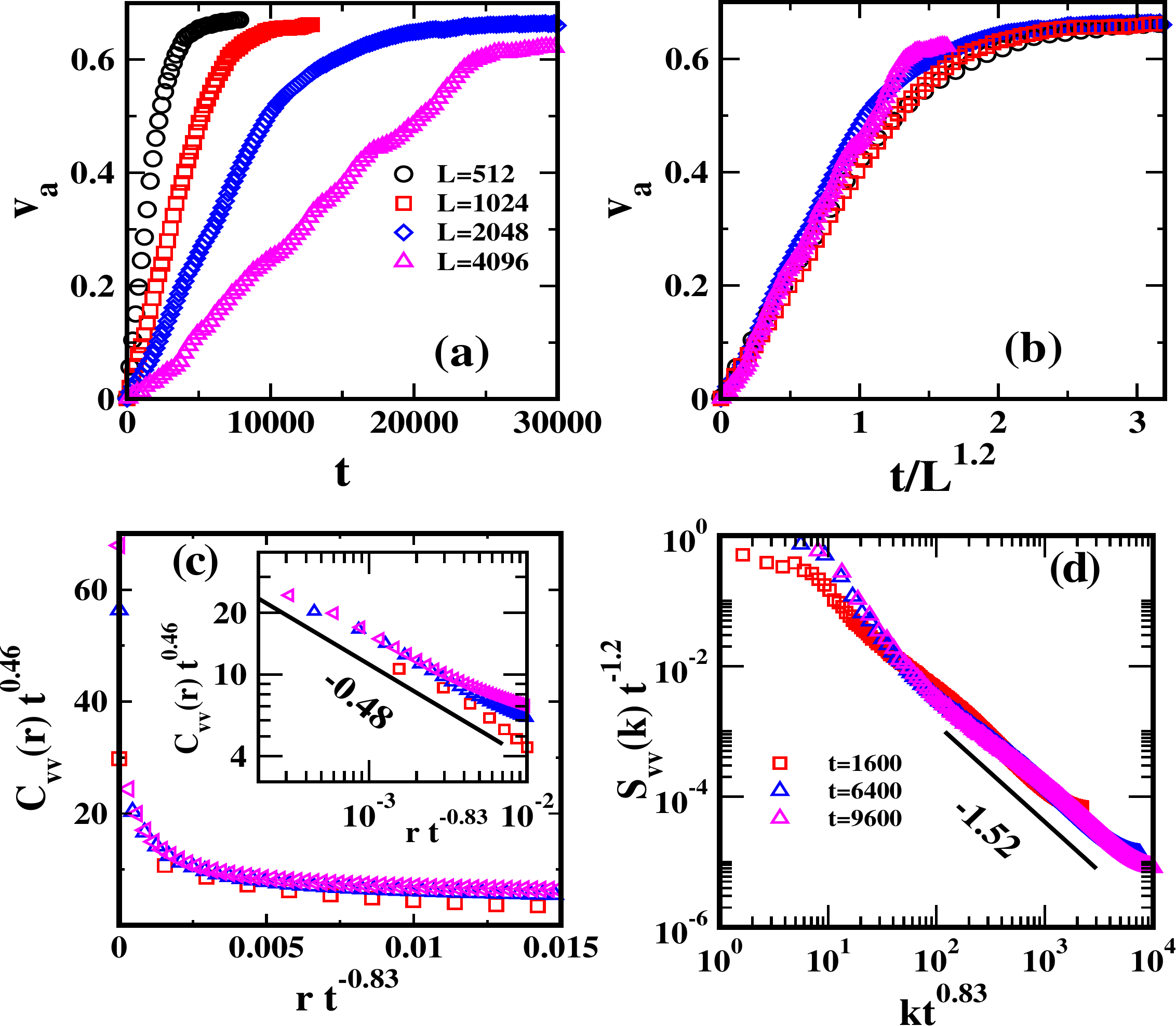}
\caption{(a) Plot of  $v_{a}$ vs. $t$ for system sizes $L=512, 1024, 2048,$ and $4096$. (b)  A data collapse of the different curves obtained as a function of scaled time $t \sim L^{z}$, with $z = 1.2$. (c) The plot of $t^{\gamma} C_{vv}(r,t)$ vs. $r/L_v(t)$ with $\gamma \simeq 0.46$ and $L_v(t) \sim t^{0.83}$ for times $t=1600, 6400$ and $9600$.  Inset: Shows log-log plot of the same data to indicate the power-law divergence with an exponent $\nu=0.48$.  (d) The scaled $S_{vv}(k,t)$ decays as a power-law with exponent $-1.52$.} 
\label{fig3}
\end{figure}

\subsection{Ordering of Velocity Field}
\label{s42}

Next we turn to the growth of the velocity order in the system. The velocity order parameter is defined as the absolute value of the sum of all particle velocities divided by $N v_0$:
\begin{equation}
{v}_{a}=\frac{1}{N v_{0}}\left |\sum_{i=1}^{N}{\bf v}_{i} \right|.
\end{equation}
The system breaks symmetry from an initially disordered state ($v_{a}=0$) to an ordered state ($v_{a}\ne 0$), such that the whole flock on an average orients towards a particular direction.  The variation of $v_{a}$ as a function of time $t$ prior to saturation is shown in Fig.~\ref{fig3}(a) for different system sizes. Note the saturation value is $\approx 0.65$.  We observe that the times $t_*$ required to attain the saturation value (in steady state) increase with the system size $L$. If we scale time as $t/L^z$ using the dynamical exponent $z = 6/5$ known from continuum theory  \cite{toner2005}, we see a good collapse of the data indicating that $t_* \sim L^{1.2}$ (see Fig.~\ref{fig3}(b)). Now at long enough times ($t \rightarrow t_*$), the coarsening length $L_v$ associated with velocity, is expected to scale as the system size $L$. This  suggests that the velocity coarsening length scale maybe $L_v(t)\sim t^{5/6}$. We test this directly as follows.

In Fig.~\ref{fig3}(c), we show the scaled correlation function $C_{vv}(r,t)$ as a function of scaled distance $r/L_v$, with a coarsening length $L_v(t)\sim t^{0.83}$. There is a data collapse for the following functional form:
\begin{equation}
C_{vv} \sim t^{-\gamma} g_v(r/L_v),
\label{eqn:vel_scaling}
\end{equation}
with $\gamma\simeq0.46$. Although not typical (with $\gamma = 0$ as in Eq.~\ref{Cr}), such cases of $\gamma \neq 0$ are known to arise in other systems  \cite{shinde}. Moreover, unlike the density correlation function (see Fig.~\ref{fig2}(a)) which has a short distant cusp singularity, we see from the inset of Fig.~\ref{fig3}(a) that $g_v(x)$ has a {\it power law divergence} at small $x$, i.e. $g_v(x)\sim x^{-\nu}$ and $\nu\simeq0.48$. Such divergence at small scaled distance is known to arise due to acute short distance ordering of non-interacting point particles, in other non-equilibrium systems undergoing density ordering \cite{shinde,apoorva05}.

\begin{figure}[!ht]
\centering
\includegraphics[width=\textwidth]{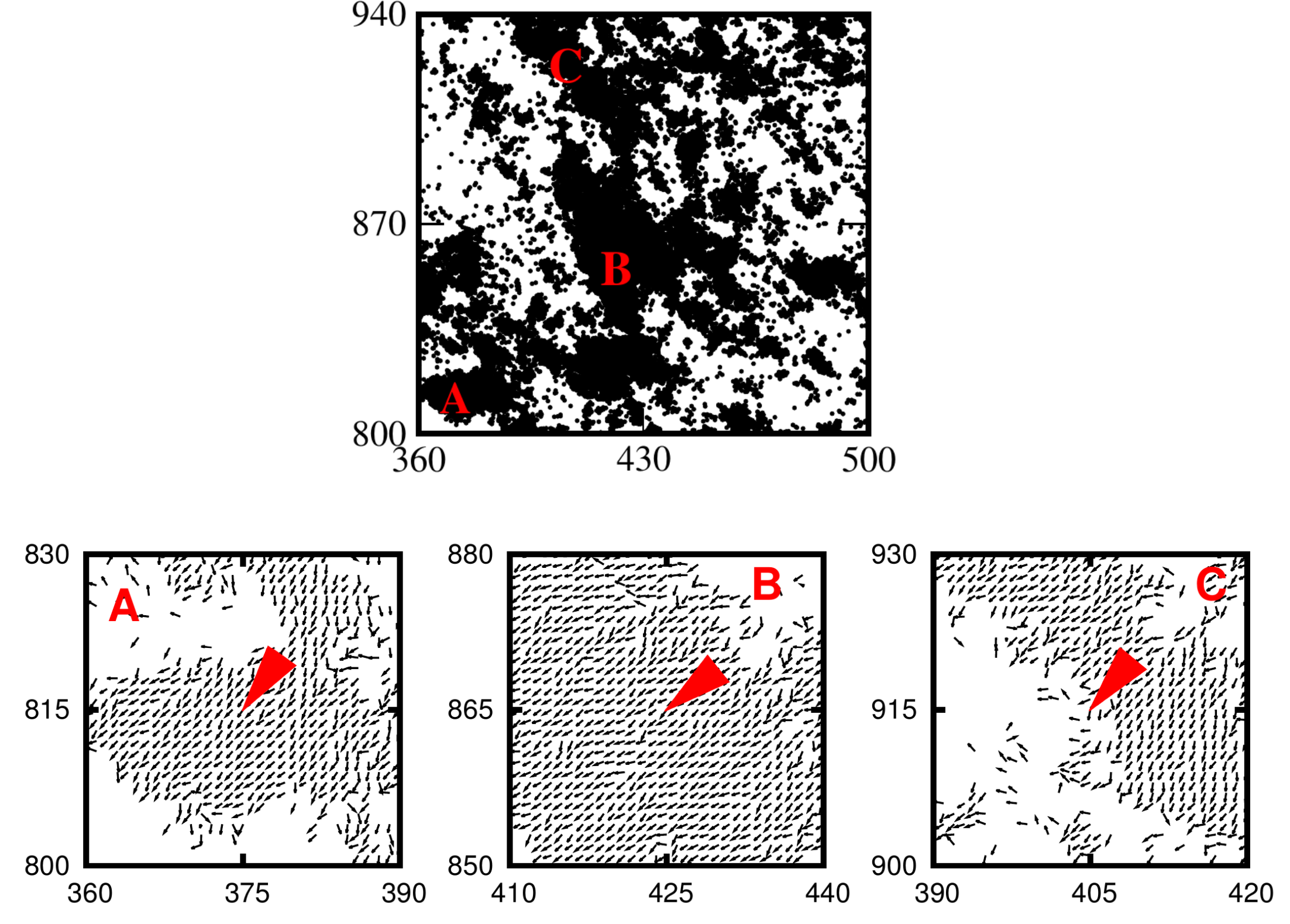}
\caption{\emph{Top}: Snapshot of a part of the system (of size $L= 1024$, at $t=5000$) showing the density clusters. We look at regions marked A, B, and C which are not part of a single compact cluster but joined by narrow tracks of particles. \emph{Bottom}: Snapshots of coarse-grained local velocities (over box size $b=1$), marked by black arrows, are shown for the three regions. The average velocity direction for each of the three regions (A, B, and C) are indicated by the red arrow heads --- those are almost perfectly aligned, showing that velocity order extends over longer distances than typical density cluster sizes.}
\label{fig4}
\end{figure}

The corresponding scaled structure factor $L_v^{-1} t^{\gamma}S_{vv}(k,t)$ against $L_v(t)$ is shown in Fig.~\ref{fig3}(b)--- a data collapse is seen for large and intermediate $kL_v(t)$, with a power law extending over almost two and half decades. The power law decay at large $kL_v(t)$ has the exponent $-(d-\nu) = -1.52$ consistent with the exponent $\nu\simeq0.48$ in the data in Fig.~\ref{fig3}(a). We observe that the velocities are after all carried by the particles. Due to the fractal morphology of the particle density clusters (discussed in the previous section), the velocity field too inherits a strong Porod law violation. We would like to stress an important point here. The velocity field does not seem to have vortices/anti-vortices (see Fig.~\ref{fig4} lower panel), which is usually responsible for BPT tails seen in vector order parameters. In the VM it seems therefore that velocity structure factors are characterised by scattering off interfacial defects, just like their density counterpart. Thus, in a sense velocity fields behave like a scalar order parameter.

The above results suggest that the growth laws for the density and velocity ordering are different. Why does the velocity field, order faster than the density? To get a qualitative idea about what is going on, we look at a locally zoomed in configuration of density clusters and the particle velocities inside those, in Fig.~\ref{fig4}. A patchy and incomplete density order (i.e. fairly disjoined clusters at best connected by some thin particle tracks) is shown in the top panel. Picking up three of such disjoined  neighbouring density clusters ($A,B$ and $C$), we check the intra-cluster coarse-grained velocities.  Those are shown in the lower panel -- the individual box velocities are marked with black arrows, while the big red arrow denote the average over a cluster.  The striking fact is that the red arrows of $A$, $B$ and $C$ are almost totally aligned.  We have checked that such a feature seen in local configurations are quite generic spatially, across the system. These observations indicate that given the same time, the velocity alignment happens over longer length scales compared to the typical density cluster sizes.

\section{Summary and Discussion}
\label{s5}

In this paper, we have studied the coarsening dynamics of the density and velocity fields of the original Vicsek model \cite{vic95}.  Unlike generic coarsening systems which typically have a  single dominant length scale, the Vicsek model quite remarkably exhibits multiple distinct coarsening length scales. The density correlations (at short and intermediate ranges) spread over a length scale $L_{\rho}(t) \sim t^{\theta_\rho}$ (with $\theta_\rho \simeq0.25$), which has been shown to be tied directly to the growing sizes of density clusters. On the other hand, the velocity correlations indicate a totally different growing length scale  $L_v(t) \sim t^{\theta_v}$ (with $\theta_v \simeq 0.83$).

In spite of the density and velocity fields being fully coupled, there is an unexpected independent dynamical evolution. The velocity order spreads over space much quicker in time, than the particles actually coming together to form density clusters. We suggest the following explanation.  Although two neighbouring density clusters may be disjoined or at best have a thin particle track connecting them, individual particles may move between clusters quite frequently due to their high speed ($v_0 = 0.5$), thereby transmitting and sharing (via local alignment rule) the orientation order over length scales bigger than sizes of the density clusters. This is possibly how velocity order spreads faster than density order. 

The fractal morphology of the density field, with two power laws ($-1.2$ and $-2.6$) charactering the structure factor at short- and long- (scaled) wave-vectors, is generally expected \cite{das} to be associated with many small sized clusters in each others neighbourhood. The latter spatial arrangement of small separations perhaps facilitates inter-cluster particle transfer more effectively and lead to a quick build up of polar order.  Finally, since the velocity field is embedded on the underlying density pattern, and we have seen that they do not have any vortex-antivortex defects, their spatial structure inherits the legacy of the underlying fractal morphology of the density. The velocity structure factor shows Porod law violation (with a decay exponent $-1.52$ at large $kL_v(t)$). \\
\ \\
{\bf Acknowledgements}: DD and SD would like to thank M. Barma, R. Rajesh and S. Mishra for useful discussions. 

%\newpage

%\newpage

\end{document}